\documentclass[preprint,showpacs]{revtex4}
\usepackage{graphicx}%
\usepackage{dcolumn}
\usepackage{amsmath}
\usepackage{latexsym}

\begin{document}
\title{ Jamming dynamics in grain mixtures : An extended hydrodynamic
approach }
\author{Supurna Sinha}
\affiliation{Raman Research Institute,
Bangalore 560 080,India}
\date{\today}
\begin{abstract}
We study jamming in granular mixtures from the novel point of
view of extended hydrodynamics. Using a hard sphere binary mixture
model we predict that a few large grains
are expected to get caged more effectively
in a matrix of small grains compared to a few small grains in
a matrix of larger ones. A similar effect has been experimentally seen in
the
context of colloidal mixtures.
\end{abstract}
\maketitle

\section{Introduction}
In recent years granular
matter has emerged as an active area of research\cite{jam}.
Interest in this field has grown as a result of observations
coming from a large number of interesting and relatively low-tech 
experiments \cite{nar}.
In particular, a vibrated granular system\cite{vib} consisting of a large 
number of 
macroscopic
grains in motion, provides us with 
a ``scaled up'' fluid system where we can explore the similarities and 
differences between such a large scale system and a microscopic atomic 
fluid. 

Recent studies\cite{jam,nagel,jamseg} in dense granular systems 
indicate that there is a striking similarity 
between the dynamics of granular materials and glassy dynamics in 
atomic fluids and colloids\cite{glass}. In this Letter, we exploit this 
analogy 
to understand some facets of jamming in granular matter. 
In particular, we study the role of compositional disorder in forming
jammed configurations which slow down the dynamics and eventually 
result in a state of structural arrest characteristic of a 
glass\cite{mcm,hardas,bossea,bosseb}. 
The observations made here are analogous to jamming 
effects studied in 
atomic liquid mixtures and colloids \cite{mcm,weitz,dsf}.  

The behavior of a fluid at large length and time scales is well
described by hydrodynamics. The hydrodynamic description has been 
extended to molecular length scales in an extended or generalized 
hydrodynamic description \cite{mcm,trk}. 
In a generalized hydrodynamic framework the basic structure of the  
hydrodynamic equations is retained and the static susceptibilities
and transport coefficients are wave-vector dependent to account for 
the nontrivial static correlations that come into play on molecular 
scales in a dense liquid.   
The generalized hydrodynamic description
has proved to be very successful in describing the dramatic 
narrowing of the central diffusive peak in a neutron scattering spectrum
$S(k,\omega)$ of a dense liquid\cite{degen}. Such a slow decay of density 
fluctuations
is a precursor to glassy dynamics in dense liquids. 
A binary liquid mixture turns out to be more effective in forming glasses 
compared to a one component liquid since compositional disorder leads to 
jammed configurations which prevent the system from reaching its
global equilibrium crystalline configuration\cite{barrat}. 
Generalized hydrodynamic studies of dense binary hard sphere mixtures in 
the 
context of glass transition suggest that at intermediate wavevectors
(i.e. equivalently on length scales of the order of the average of the  
molecular diameters of the two species) 
the density fluctuations of the two 
species emerge as the slowest
decaying fluctuations and therefore dominate the slow dynamics of the 
system \cite{mcm,dsf}. 
In a self-consistent mode-coupling theory 
(MCT), nonlinear 
couplings of these slowly decaying modes of
density fluctuations lead to a glassy state where structural relaxation 
is frozen\cite{hardas,bossea}.   
  
In this Letter we use, for the first time, 
generalized hydrodynamic 
techniques developed in 
the realm of atomic
liquids to understand the behavior of jammed configurations in granular 
mixtures. 
Such an approach to granular matter enables us to take into consideration 
non-trivial static correlations stemming from the {\it granularity} or
{\it finite sizes} of the particles, which in turn, influences the high
density jamming dynamics of the system. 
We thus gain new insight into granular mixture 
dynamics
and make predictions for future experiments in such systems.  

\section{Extended Hydrodynamic Approach To A Binary Mixture Of Grains}

Consider a binary granular mixture of hard spheres of diameters 
$\sigma_1$
and $\sigma_2$ ($\sigma_2$$>$$\sigma_1$), 
masses $m_1$ and $m_2$,
number densities $n_1$
and $n_2$ and of total packing fraction $\eta = \frac{\pi}{6}
[n_1 \sigma_1^{3}+n_2 \sigma_2^{3}]$.  
An analysis of the extended hydrodynamic equations of such 
a binary hard sphere mixture suggests that on length scales 
of the order of the size of the hard sphere particles
\cite{mcm} momentum and temperature
fluctuations decay very fast and the slow dynamics of the system 
can be well described in terms of the modes of density fluctuations of 
the two species\cite{dsf,foot}. 
This enables us to confine to a two mode description 
of the system on length scales 
of the order of the diameters of the hard sphere granular particles
at high densities. 
It is convenient to describe the system in terms
of the following linear combinations of the density fluctuations
of the two species-
the total mass density fluctuation at wavevector ${\vec k}$
$$\rho_{\vec k} = \rho_{1\vec k}+\rho_{2\vec k}$$
and the concentration fluctuation at wavevector ${\vec k}$
$$c_{\vec k} = \frac{\rho_{2}}{\rho^{2}}\rho_{1\vec 
k}-\frac{\rho_{1}}{\rho^{2}}\rho_{2\vec k}$$
where $\rho_{1\vec k}$ and $\rho_{2\vec k}$ are the mass density 
fluctuations of species $1$ and $2$, $\rho_1 = m_1 n_1$ and 
$\rho_2 = m_2 n_2$ are the equilibrium mass densities of species
$1$ and $2$ and $\rho = \rho_1 + \rho_2 $ is the total equilibrium
mass density. 
The set of Laplace transformed coupled extended hydrodynamic equations of 
this system is given by:
\begin{equation}
\left[z+\frac{k^2}{\rho \chi_T(k)\gamma_L(k)}\right]{\rho}_{\vec k}(z)\\
+\frac{k^2 \rho}{\beta \gamma_L(k)}\\
\left[\frac{f_1(k)}{m_1}-\frac{f_2(k)}{m_2}\right]{c}_{\vec k}(z) \\
={\rho}_{\vec k}(t=0)
\label{lapa}
\end{equation} 
and 
$$\left[z+k^2{D(k)}\left[\frac{m_2}{m_1}{\sqrt{\frac{n_2}{n_1}}}f_1(k)
+\frac{m_1}{m_2}{\sqrt{\frac{n_1}{n_2}}}f_2(k)\right]\right]{c}_{\vec 
k}(z)$$
\begin{equation}
+\frac{k^2}{{\rho}^2}D(k)\sqrt{n_1n_2}\\
\left[{m_1}{f_2(k)}-{m_2}{f_1(k)}\right]{\rho}_{\vec k}(z) \\
={c}_{\vec k}(t=0)
\label{lapb}
\end{equation} 
where $\chi_T(k)$ is the generalized isothermal compressibility,
defined in terms of the partial static structure factors
$S_{ij}(k)$ with $i=1,2$ and $j=1,2$:
$$ \chi_T(k)=\chi_T^{0}\frac{S_{11}(k)S_{22}(k)
-S_{12}^{2}(k)}
{x_2 S_{11}(k)+x_1 S_{22}(k)-2\sqrt{x_1 x_2}S_{12}(k)}$$
Here $\chi_T^{0}$ is the 
compressibility of the granular gas in the dilute limit. 
$\gamma_L(k)$ is the generalized 
longitudinal viscosity\cite{mcm}.  
$$f_1(k)=\frac{S_{22}(k)+\frac{m_1}{m_2}\sqrt{\frac{n_1}{n_2}}S_{12}(k)}
{S_{11}(k)S_{22}(k)-S_{12}^{2}(k)}$$ 
and 
$$f_2(k)=\frac{S_{11}(k)+\frac{m_2}{m_1}\sqrt{\frac{n_2}{n_1}}S_{12}(k)}
{S_{11}(k)S_{22}(k)-S_{12}^{2}(k)}$$ 
are combinations of partial static structure factors and
$D(k)$ is the coefficient of mutual diffusion\cite{mcm,ferz}. 

This set of equations leads to two extended hydrodynamic diffusive modes.

Since we are interested in exploring the packing aspects of jamming
in a binary granular mixture
which is controlled by the {\it sizes} rather than the masses of the 
particles, we confine ourselves to the case of equal masses $m_1=m_2$
for the two species\cite{barrat}.
Here we analyze two illuminating special cases to bring out
the role of size difference and packing in the jamming process which 
triggers the transition to a glassy state:
$(i)$ a system composed of a few large spheres in a matrix of small 
spheres and  
$(ii)$ a system composed of a few small spheres in a matrix of large 
spheres.
 
In both these extreme limits ($x_2=n_2/n <<1$ and $x_1=n_1/n <<1$)
the cross terms in the expressions representing the eigenvalues
for the extended diffusive modes
are negligibly 
small and the mode 
structure 
reduces to:

\begin{equation}
z_{-}(k)\\
\simeq -{k^2}\frac{1}{\rho \chi_T(k)\gamma_L(k)}\\
\label{minus} 
\end{equation}

and

\begin{equation}
z_{+}(k)\\
\simeq -{k^2}D(k)[\sqrt\frac{n_2}{n_1}f_1(k)\\
+\sqrt\frac{n_1}{n_2}f_2(k)] \\ 
\label{plus}
\end{equation}

Thus, there are two relevant modes: $z_{-}(k)$, which governs the 
relaxation of total mass density fluctuations and $z_{+}(k)$,
which governs the relaxation of concentration fluctuations. Let us
analyze these modes for case (i). In this case, since the mixture
consists of a large number of small spheres, the static compressibility
$\chi_T(k)$ which is the main determinant of the dynamics of a dense 
liquid, is given by $\chi_T(k) \simeq \chi_T^{0} S_{11}(k)$,
i.e. it is dominated by the static structure factor of the small spheres. 
Thus, $z_{-}(k)\simeq -{k^2}\frac{1}{\rho \chi_T^{0} S_{11}(k)\gamma_L(k)}$.
Consequently there is a significant slowing down of the dynamics of 
density fluctuations of the background matrix at the location of the
first peak of $S_{11}(k)$\cite{footnote}. In this case the static 
structure of large (type $2$) spheres is flat 
and is given by $S_{22}(k) \simeq 1$.
Thus, $z_{+}(k)\simeq -{k^2}\frac{D_{02}}{ S_{22}(k)}
\simeq -{k^2}{D_{02}}$, where $D_{02}$ is the dilute gas limit of the 
diffusion coefficient of large spheres\cite{ferz}. 
In case (ii), the large (type $2$) and small (type $1$)
spheres switch roles. The final picture that emerges is the 
following. Caging and slowing down of dynamics is more effective for 
$x_2<<1$ firstly because the ratio $D_{02}/ D_{01}=\sigma_1/\sigma_2 <1$
\cite{mcm,vijay} 
and a few large spheres
diffuse {\it slower} in the background of small spheres compared to 
a few small spheres in the background of large spheres. 
In addition, in both cases there is also a significant slowing 
down of structural relaxation due to softening of the mode of
total density fluctuations $z_{-}(k)$ 
around the location of the peak 
of the static structure factor of the majority particles.

In other words, caging is more efficient in a mixture dominated by 
small spheres compared to one dominated by large spheres. 
This is the main prediction made in this Letter for 
future experiments on dense granular mixtures designed to probe 
the efficiency of caging in such systems.   
 
The main point that we emphasize in this Letter is that by exploiting 
the analogy between glass forming atomic liquids and granular
matter we can draw some definite testable conclusions about 
caging dynamics in grains. 

To summarize, we have, for the first time applied extended or generalized
hydrodynamic techniques to dynamics of granular matter.
In particular,  
we consider a dense binary mixture of hard spheres and 
analyze the modes of density fluctuations which dominate the slow 
dynamics on length scales of the order of the sizes of the 
granular particles constituting the system. Our analysis points to some 
differences in 
jamming behavior between a mixture dominated by small spheres and
one dominated by large spheres. 
Effects similar to the ones predicted here have been observed in 
confocal microscopy studies in colloidal mixtures \cite{weitz}.
The predictions made here can be 
tested against experiments in vibrated dense granular mixtures. 
We expect our analysis to be valid for a granular mixture consisting of 
nearly elastic spheres of comparable masses \cite{duf}. 
While the present analysis captures the onset of glassy behavior in 
granular mixtures, it would be worthwhile to do a mode-coupling 
study for such a system using the extended hydrodynamic modes
that stem out of our analysis as an input to understand glass transition
in granular mixtures. 
In future, one can do a more complete 
analysis where
the effects of momentum and temperature fluctuations and inelastic
collisions \cite{brey} are taken into consideration and 
check if the 
high density particle-scale dynamics presented here survives the 
inclusion of these effects.


\begin{thebibliography}{1}
\bibitem{jam} See for instance, H. A. Maske, J. Brujic and  S. F. Edwards,
{\it The physics of Granular Media}, Wiley-VCH, (2004).
\bibitem{nar} See for instance,
K. Feitosa and N. Menon, {\it Physical Review Letters} {\bf 88 },
198301 (2002).
\bibitem{vib} X. Yang, C. Huan, D. Candela, R. W. Mair and R. L. Walsworth
{\it Phys. Rev. Lett.} {\bf 88}, 044301 (2002);
X. Yang, C. Huan, D. Candela, R. W. Mair and R. L. Walsworth
{\it Physical Review E} {\bf 69}, 041302 (2004);
\bibitem{nagel} A. J. Liu and S. R. Nagel, {\it Nature} {\bf 396}, 21
(1998).
\bibitem{jamseg} See for instance M. Nicodemi et al,
``Statistical Mechanics of jamming and segregation in granular media''
in ``Unifying Concepts in Granular Media and Glasses'' edts. A. Coniglio,
A. Fierro, H. J. Herrmann and M. Nicodemi.
\bibitem{glass} W. Gotze and L. Sjorgen, {\it Rep. Prog. Phys} {\bf 55},
241 (1992).
\bibitem{mcm}
M. C. Marchetti and Supurna Sinha,
{\it Physical Review A } {\bf 41} 3214 (1990);
Supurna Sinha and M. C. Marchetti,
{\it Physical Review A} {\bf 46} 4942 (1992).
\bibitem{hardas}
U. Harbola and S. P. Das
{\it Physical Review E} {\bf 65}, 036138 (2002).
\bibitem{bossea} J. Bosse and J. S. Thakur, {\it Physical Review Letters}
{\bf 59}, 998 (1987).
\bibitem{bosseb} J. Bosse and Y. Kaneko, {\it Physical Review Letters}
{\bf 74}, 4023 (1995).
\bibitem{weitz}
A. D. Dinsmore, E. R. Weeks, V. Prasad, A. Levitt and D. A. Weitz,
{\it Applied Optics} {\bf 40}, 4152 (2001).
\bibitem{dsf}
Supurna Sinha
{\it Physical Review E} {\bf 49} 3504 (1994).
\bibitem{trk}
T. R. Kirkpatrick,
{\it Physical Review A} {\bf 32}, 3130 (1985).
\bibitem{degen}
P. G. de Gennes,
{\it Physica} {\bf 25}, 825 (1959).
\bibitem{barrat}
J. N. Roux, J. L. Barrat and J. P. Hansen
{\it Journal Of Physics : Condensed Matter} {\bf 1}, 7171 (1989).
\bibitem{foot}
A more familiar example is the phenomenon of de Gennes narrowing which
happens in the context of dense simple liquids.
\bibitem{ferz}
J. Ferziger and H. Kaper,
{\it Mathematical Theory Of Transport Processes in Gases},
{North-Holland, Amsterdam,} (1972).
\bibitem{footnote} Also, we are focussing on a high viscosity
(i.e. large $\gamma_L(k)$[\cite{mcm,trk}]) regime which contributes
to the process of structural slowing down.
\bibitem{vijay}
See for instance,
G. V. Vijayadamodar and B. Bagchi
{\it Journal Of Chemical Physics} {\bf 93}, 689 (1990).
\bibitem{duf} In such a domain the mixture can be described fairly
well as one sharing a common temperature $T$. See for instance
P. Zamankhan {\it Physical Review E} {\bf 52}, 4877 (1995)
and
V. Garzo and J. W. Dufty {\it Physical Review E} {\bf 60}, 5706 (1999).
\bibitem{brey} See, for instance,
J. W. Dufty and J. J. Brey {\it Physical Review E} {\bf 68}, 030302 
(2003); 
J. J. Brey and J. W. Dufty {\it Physical Review E} {\bf 72}, 011303 
(2005); 
\end{thebibliography}
\end{document}